\documentclass[reprint,amsmath,amssymb,aps]{revtex4-1}
\usepackage{graphicx}
\usepackage{dcolumn}
\usepackage{bm}
\usepackage{xcolor}

\usepackage{graphicx}
\begin{document}

\preprint{APS/123-QED}

\title{Explosive death in nonlinear oscillators coupled by quorum sensing}

\author{Umesh Kumar Verma, Sudhanshu Shekhar Chaurasia and Sudeshna Sinha}
\affiliation{Indian Institute of Science Education and Research Mohali, Knowledge City, SAS Nagar, Sector 81, Manauli, Punjab, PO 140 306, India}

\date{\today}

\begin{abstract}
Many biological and chemical systems exhibit collective behavior in response to the change in their population density. These elements or cells communicate with each other via dynamical agents or signaling molecules. In this work, we explore the dynamics of nonlinear oscillators, specifically Stuart-Landau oscillators and Rayleigh oscillators, interacting globally through dynamical agents in the surrounding environment modeled as a quorum sensing interaction. The system exhibits the typical continuous second-order transition from oscillatory state to death state, when the oscillation amplitude is small. However, interestingly, when the amplitude of oscillations is large we find that the system shows an abrupt transition from oscillatory to death state, a transition termed ``explosive death". So the quorum-sensing form of interaction can induce the usual second-order transition, as well as sudden first-order transitions.  Further in case of the explosive death transitions, the oscillatory state and the death state coexist over a range of coupling strengths near the transition point. This emergent regime of hysteresis widens with increasing strength of the mean-field feedback, and is relevant to  hysteresis that is widely observed in biological, chemical and physical processes.
\end{abstract}

\pacs{Valid PACS appear here}  
\maketitle
\section{Introduction}

The dynamics of many complex real-world systems can be described as an ensemble of coupled oscillators and the study of the collective behavior of such systems  has attracted a lot of research attention~\cite{watts}.  The interaction among the dynamical systems can lead to interesting emergent phenomena such as synchronization, amplitude modulation,  phase flip, and suppression of  oscillation. Among these, synchronization is one of the most widely studied field and synchronization transitions are often used to describe the changes from incoherence to coherence in large interactive dynamical systems~\cite{sync1}. It  has relevance in contexts ranging from neuronal networks to communication and laser systems~\cite{Pikovsky01}.  Suppression of oscillations in coupled systems is another emergent phenomenon of considerable relevance in such complex systems. Depending on the nature of the emergent steady states and the underlying mechanisms that create these states, suppression of oscillations falls under two classes: Amplitude death(AD) and Oscillation death (OD). In AD, coupled oscillators arrive at a common stable steady state, which is the fixed point of the systems~\cite{AD}.  However, in case of OD, coupled oscillators go to a new coupling-dependent steady state(s). In this case, the oscillators may go to different steady states (termed inhomogeneous steady states, IHSS) arising from a symmetry breaking bifurcation, or the oscillators may all go to a coupling-dependent homogeneous steady state (HSS)~\cite{OD,revival1,revival2,OD_pre}. 

In networked oscillators, it is essential to investigate how the transitions from incoherence to coherence, or oscillatory behaviour to death states occur. An important question here is if the transition process is continuous or discontinuous (abrupt).  A classical result is that in most of the cases these transitions
are continuous and reversible, i.e., a second-order type. However, a discontinuous and irreversible (i.e., a first-order type) transition from incoherence to coherence state (called Explosive Synchronization (ES)) was also found for Kuramoto model on a scale-free network when the natural frequency and degree of nodes were correlated ~\cite{ES1}. Since then, ES has been extensively studied in various network models~\cite{ES2}, but it has been found only in phase synchronization, either in the Kuramoto-type model or in the coupled R\"{o}ssler system. Apart from ES, recently,  a first-order like discontinuous and irreversible transition from the oscillatory state to the death state, termed Explosive Death (ED), has been found in a frequency-weighted Stuart-Landau oscillator network model~\cite{ED1}. Subsequently, ED has been found in both mean-field coupled limit cycle and chaotic oscillators under varying intensity of the mean–field interaction~\cite{ED2}. The ED phenomenon has been also observed in indirect coupling, where oscillators communicate with a common environment~\cite{ED3} and conjugate coupling, where oscillators interact through interaction with a dissimilar variable~\cite{ED4}. As in case of the ED transition, near the transition point, two states (namely, the oscillatory state and the death state) coexist over a range of coupling strengths. The co-existence of oscillatory states and death states has been observed in many physical systems~\cite{experi-phy1,experi-phy2}, chemical systems~\cite{experi-chem1,experi-chem1},  and also in various models of limit cycle and chaotic systems~\cite{coexistence1,coexistence12}.

Now, the dynamics of many biological and chemical systems arise from collective behavior of a large number of dynamical units coupled via local dynamical agents or signaling molecules. In biological systems, such as suspensions of yeast in nutrient solutions~\cite{yest}, starving cellular colonies of the social amoeba Dictyostelium discoideum~\cite{bio1}, and synthetically engineered bacteria~\cite{bio2}, the cell population density is an important key for the collective behavior of the cell.  A well-known example of emergence of collective behavior in bacteria is the {\em quorum sensing mechanism}, where change in the population density of the elements above a threshold value triggers a sudden transition from a quiescent state to synchronized oscillations~\cite{bio3}. In a neuronal context, a mechanism similar to that of quorum sensing may involve local–field potentials, which may play an important role in the synchronization of groups of neurons~\cite{bio4}.  A similar quorum sensing-type phenomenon also appears in the Belousov-Zhabotinsky (BZ)-reaction, where cation-exchange beads in the solution also depend on the number density of the beads~\cite{chem1,chem2}. Other examples of such a mechanism are cold atoms interacting with a coherent electromagnetic field~\cite{cold}, and the onset of coordinated activity in a population of microorganisms living in a shared environment~\cite{micro}. 

The significance of quorum sensing in the collective dynamics of coupled dynamical systems, in particular many biological and chemical systems, has made it a focus of intense research activity~\cite{s1,s2,s3,s4,s5,s6,s7,s8,s9}.  However, the emergent collective behaviour of nonlinear oscillators interacting through dynamical agents, has not been studied from the point of view of {\em phase transitions}. So in this work we focus on the transitions in populations of limit-cycle oscillators interacting through dynamic agents/signaling molecules in the surrounding environment, and analyze these transitions in the framework of phase transitions. In our coupling scheme the oscillators interact with each other through its dynamical agents, which in turn interact globally with each other in the surrounding medium. Interestingly, we will show that this type of interaction yields a first-order transition from oscillatory state to death state and vice versa, as coupling strength is varied.

The paper is organized as follows: In Section II, we describe the model system of nonlinear oscillators, interacting through dynamic agents in the surrounding environment. In Sections III and IV, we demonstrate the Explosive Death transition through order parameter continuation diagrams, bifurcation analysis, phase diagrams in the parameter plane of the most relevant parameters and time series analysis, for the cases of coupled Stuart-Landau  oscillators and coupled Rayleigh oscillators respectively. Finally, in Section V, we present a summary and discussion of our salient results.

\section{Model of Nonlinear Oscillators Coupled via Quorum Sensing Interaction}
We consider $N$  nonlinear oscillators interacting through dynamic agents in the surrounding environment~\cite{bio3,s10}. The dynamics of such coupled systems can be modelled as:
\begin{eqnarray}
\dot{\bm{X}}_i &=& \bm{F}(\bm{X}_i) +\varepsilon \bm{\kappa} s_i \nonumber \\
\dot{s}_i &=& -\gamma_i s_i-\varepsilon {\bm{\kappa}^{T} X_i}+\eta(Q\bar{s}-s_i) 
\label{eq1}
\end{eqnarray}

where $i=1,2,\ldots,N$. Here $\bm{X}_i= \begin{bmatrix} x^1_i & x^2_i & \ldots & x^m_i \end{bmatrix}^{T}$ represents the state variables of  $m$--dimensional nonlinear oscillators, whose dynamical equations are given by $\bm{F}(\bm{X}_i)=\begin{bmatrix} f^1_i(\bm{X}_i) &f^2_i(\bm{X}_i)& \ldots & f^m_i(\bm{X}_i) \end{bmatrix}^{T}$. The $i^{th}$ dynamical oscillator directly interacts with its dynamic agent $s_i$ with strength $\varepsilon$ and all these agents interact with each other in the surrounding environment. The dynamics of the dynamic agents $s_i$ are considered to be a one-dimensional over-damped oscillators with damping coefficient $\gamma_i>0$. The matrix $ \bm{\kappa}=\{ \kappa_i\}$ is a column matrix having dimensions $ m \times 1$, with elements $0$ and $1$, and it determines the components of $X_i$ that get feedback from the dynamic agents. The transpose of ${\bm{\kappa}}$, ${\bm{\kappa}^{T}}$, decides the components of $X_i$ which give feedback to the local agent. The strength of coupling between the dynamic agents and the oscillators $\varepsilon$, the diffusion coefficient $\eta $ and the strength of mean-field interaction of  all the dynamical agents $Q$ are the important parameters in the system.

In the above model system, local dynamical agents $s_i$ represent the particle species that can freely diffuse in the medium and allow individual oscillators to communicate with each other. The specific realization of local dynamical agents $s_i$ depends on the context. In synthetic bacteria these local dynamical agents $s_i$ are chemical signal molecules (called auto-inducers) which can freely diffuse across the cell membrane and allow the bacteria to sense a critical cell mass and respond to the activation of receptors on the cell membrane~\cite{bio3}. Similarly in the BZ reaction, $s_i$ represents the  chemical species that diffuse between autocatalytic beads~\cite{chem1,chem2}.

\section{Coupled Stuart-Landau Oscillators}

We first consider an ensemble of $N$ Stuart-Landau oscillators (SL), interacting through dynamic agents in the surrounding environment as modelled by Eqn.~\ref{eq1}. The dynamics of coupled system is then given by
\begin{eqnarray}
\dot{x}_i &=& (\rho-x_i^2-y_i^2)x_i-\omega_i y_i +\varepsilon s_i \nonumber \\
\dot{y}_i &=& (\rho-x_i^2-y_i^2)y_i+\omega_i x_i \nonumber \\
\dot{s}_i &=& -\gamma_i s_i-\varepsilon x_i+\eta(Q\bar{s}-s_i) 
\label{eq2}
\end{eqnarray}

where, $i=1,2,\ldots,N$ and $(x_i,y_i)$ are the state variables of $i^{th}$ oscillators. The individual  oscillator exhibits self-sustained limit cycle oscillations with amplitude $\sqrt{\rho}$ and natural frequency $\omega_i$.

\begin{figure}
\includegraphics[width=0.45\textwidth]{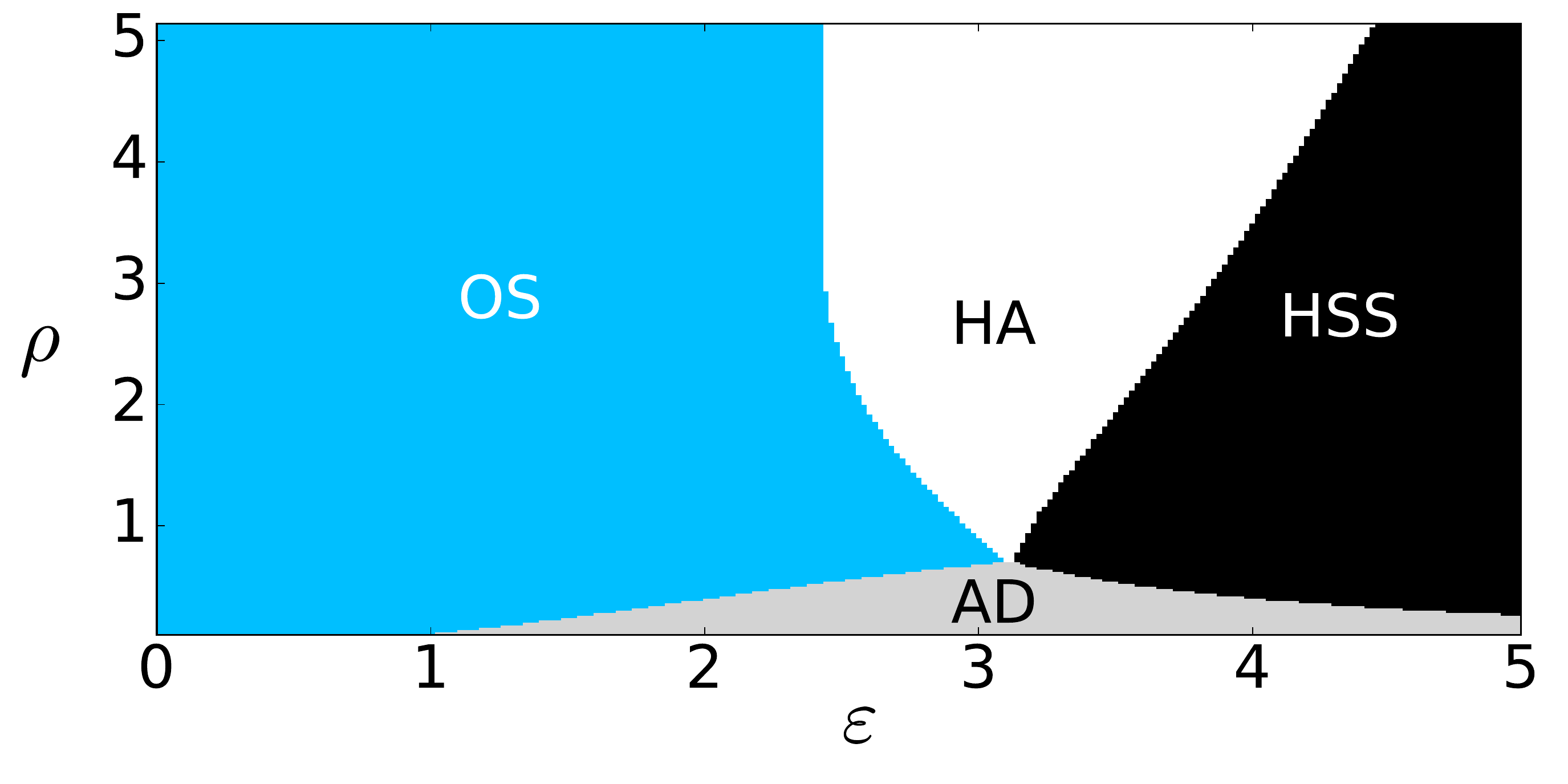}
\caption{(Color online) Different dynamical domains of $N$ coupled SL oscillators (cf. Eqn~\ref{eq2}) in the parameter plane $(\varepsilon -\rho)$. The marked regime OS, HA, HSS,and AD represent the oscillatory state, hysteresis, homogeneous steady state, and amplitude death state respectively.The other parameters are $\gamma=1$, $\omega=2$, $Q=0.5$, $\eta=1$ and $N=100$.}
\label{fig1}
\end{figure}

To monitor the explosive transition as coupling strength $\varepsilon$ grows, we adopted an order parameter $A(\varepsilon)$~\cite{ED2}, which is given by
\begin{equation}
\begin{split}
 A(\varepsilon)=\frac{a(\varepsilon)}{a(0)}.
\end{split}
\label{eq3}
\end{equation}

Where $ a(\varepsilon) = \frac{1}{N}\sum_{i=1}^N \left[\langle x_{i,max} \rangle_t - \langle x_{i,min}\rangle_t \right]$ represent the difference of the global maximum and minimum values of the time series of oscillator over a long transient stage at a particular coupling strength $\varepsilon$. Hence for oscillatory state the value of order parameter $A(\varepsilon)>0$, while for the steady state the  value of $A(\varepsilon)=0$.   

Throughout the work, the dynamics of coupled oscillators are solved using the fourth--order Runge--Kutta method. In both forward and backward continuation, order parameter $A(\varepsilon)$ is calculated by the classic adiabatic method.  In the case of forward continuation, first $A(0)$ is calculated for some random initial condition, and then the value of  $\varepsilon$ is gradually increased from $\varepsilon_0$ to $\varepsilon_{max}$, in steps of size $\delta\varepsilon$. The final state of the prior $\varepsilon$ used as the initial condition for the next $\varepsilon$. The backward continuation is also  performed in the same way by gradually decreasing the value of $\varepsilon$ from $\varepsilon_{max}$ to $\varepsilon_0$. Additionally, in order to increase the robustness of our results, we add small noise (of the order of $10^{-6}$) to the initial state of the system at each value of $\varepsilon$, as we sweep $\varepsilon$ in forward and backward continuation.
 Here we used $\delta\varepsilon = 0.02$, without loss of generality.

 To understand the effect of the amplitude of oscillations of the coupled SL systems, we plotted  a phase diagram $(\varepsilon-\rho)$ adiabatically in both forward and backward continuation of $\varepsilon$, which is shown in Fig.~\ref{fig1}. In this figure region OS, AD, HSS, and HA  denote oscillatory state , amplitude death state, homogeneous steady state,  and  hysteresis area (where both OS and HSS coexist in the parameter space) respectively. It is clearly evident in this figure that when $\rho<0.75$ the system stabilized at AD and there is no hysteresis in the system. But when $\rho>0.75$ coupled system stabilized at a coupling-dependent homogeneous steady state (HSS) and there is a hysteresis in the system where OS and HSS solution co-exist. Interestingly, we can see that an increase of $\rho$ leads to an increase in the hysteresis area in parameter space.    

\begin{figure}
\includegraphics[width=0.45\textwidth]{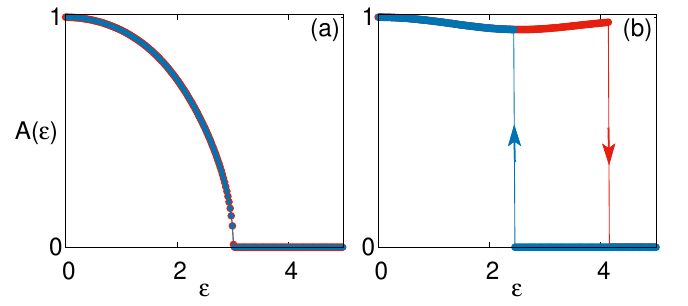}
\caption{(Color online) Forward and backward continuation of the Order Parameter, $A(\varepsilon)$, under the variation of the coupling strength $\varepsilon$, for $N$ coupled SL oscillators is shown for (a) $\rho=0.75$ and (b)  $\rho=4$. The other parameters are $\gamma=1$, $\omega=2$, $Q=0.5$, $\eta=1$ and $N=100$.}
\label{fig2}
\end{figure}

 To characterize the behavior of the transition from oscillatory state to death state or vice versa, we calculate the variation of order parameters $A(\varepsilon)$  with respect to variation of coupling strength $\varepsilon$, for both forward and backward continuation, at different values of $\rho$. In Fig.~\ref{fig2}(a), the behavior of the order parameter for $\rho=0.75$, shows a {\em continuous transition} from oscillatory state to death state and vice versa in  both forward and  backward continuation.  In this case, both forward and backward transitions occur at the same critical value of $\varepsilon$ and clearly indicates a {\em second order transition} from oscillatory state to death state. In Fig.~\ref{fig2}(b), for $\rho=4$, the variation of $A(\varepsilon)$ clearly indicates an {\em abrupt} transition from oscillatory state to death state and vice versa in both forward and backward continuation respectively. The forward and backward transitions points occur at a different values of $\varepsilon$, and the system exhibits {\em hysteresis} in the parameter space. So interestingly, this interaction yields both first-order and second-order transitions, depending on the amplitude of the constituent oscillators in the network.
 
\begin{figure}
\includegraphics[width=0.45\textwidth]{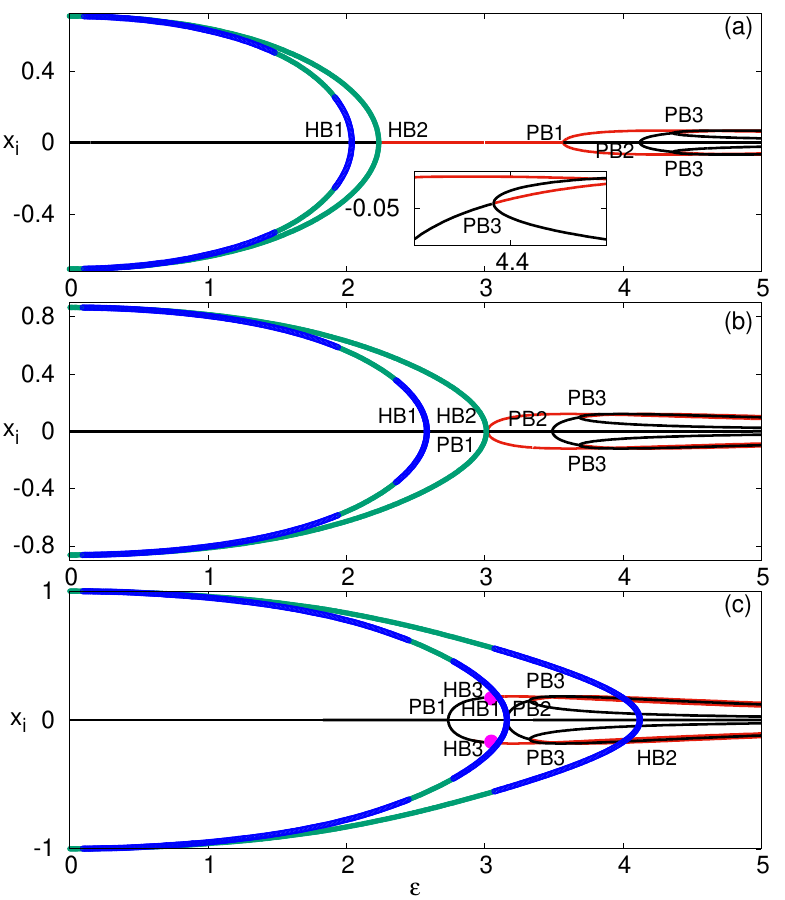}
\caption{(Color online) Bifurcation diagram of two coupled SL oscillators is plotted with respect to coupling strength $\varepsilon$  for (a) $\rho=0.5$, (b) $\rho=0.75$, and (c) $\rho=1.0$. Inset of (a) shows the zoomed in view of the  pitchfork bifurcation PB3. Here the red and black lines show the stable and unstable steady states respectively, while green and blue  circles represent the stable and unstable periodic solutions respectively. The bifurcation point namely pitchfork bifurcation (PB) and Hopf bifurcation (HB) are also marked. The other parameters are $\gamma=1$, $\omega=2$, $\eta=1$, and $Q=0.5$.}
\label{fig3}
\end{figure}

Fig.~\ref{fig3} (a) shows the bifurcation diagram of two coupled SL oscillators, obtained using the software XPPAUT~\cite{xpp}. The figure displays $x_1, x_2$  with respect to coupling strength $\varepsilon$, for  $\gamma=1$, $\omega=2$, $\eta=1$, and $Q=0.5$. Here the bifurcation points (HB1, HB2, HB3) represent Hopf bifurcation points, while bifurcation points (PB1, PB2, PB3) denote pitchfork bifurcation points. The red and black lines represent stable and unstable steady states, while the green and blue circles represent stable and unstable limit cycles respectively.  From this figure we see that the coupled system is first stabilized at the origin through Hopf bifurcation HB2. Subsequently a HSS solution is born through a pitchfork bifurcation PB1. Further on, at higher values of coupling strength $\varepsilon$, IHSS solutions are born through a pitchfork bifurcation PB2 and stabilized through another pitchfork bifurcation PB3. Now, with increasing $\rho$ the bifurcation points HB2 and PB1 come closer to each other, and at a particular $\rho$ value  ($\rho_c \sim 0.75$), HB2 collides with PB1, as seen in Fig.~\ref{fig3}(b). Further, for $\rho > \rho_c$, the bifurcation points HB1 and HB2 move to the right of PB1, and the bifurcation diagram for $\rho=1>\rho_c$ is shown in Fig.~\ref{fig3}(c). For $\rho>\rho_c$, the HSS solution born through a PB1 but stabilized through a subcritical Hopf bifurcation HB3, is marked by a point in Fig.~\ref{fig3}(c). We also found that when $\rho>\rho_c$, the transition form oscillatory state to death state is abrupt and discontinuous and Hopf bifurcation HB3 is the backward transition point of this {\em first-order transition}.  

\begin{figure}
\includegraphics[width=0.45\textwidth]{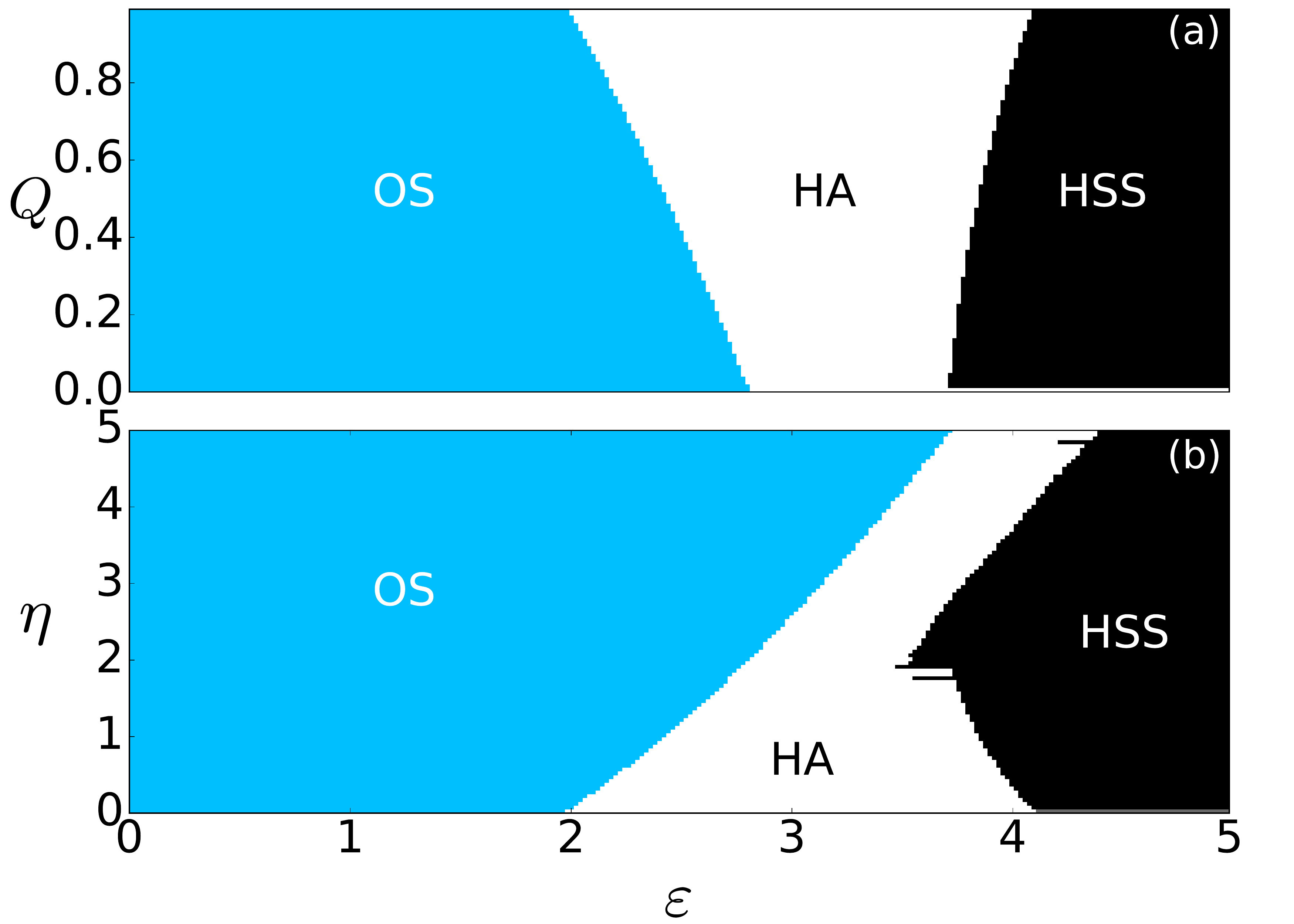}
\caption{(Color online) Different dynamical domains of $N$ coupled SL oscillators are plotted in the parameter plane (a) $(\varepsilon -Q)$ for $\eta=1$,  and (b)$(\varepsilon -\eta)$ for $Q=0.5$. The other parameters are $\rho=3.5$, $\gamma=1$, $\omega=2$ and $N=100$.}
\label{fig4}
\end{figure}

\begin{figure}
\includegraphics[width=0.45\textwidth]{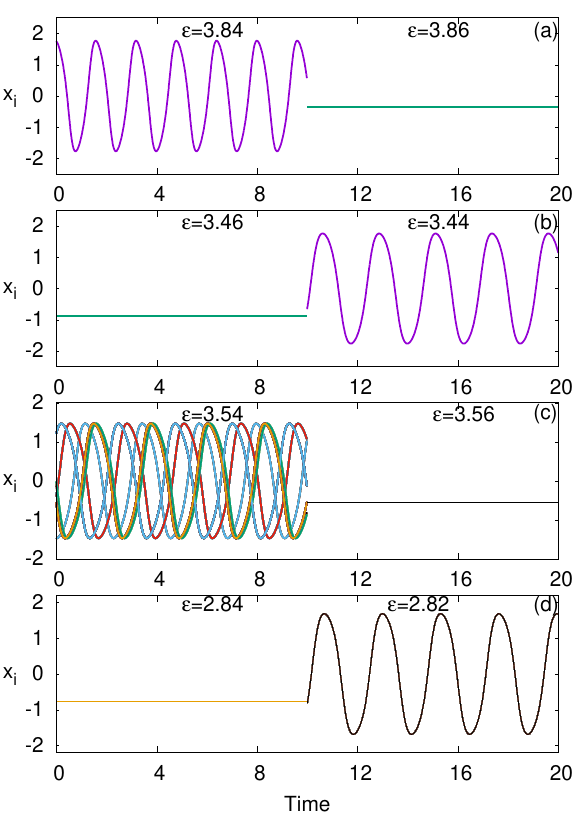}
\caption{(Color online) The time series, corresponding to Fig.~\ref{fig4}(b), in two different regimes near the transition points in both forward and backward continuation (a,b) for $\eta=1$ and (c,d) for $\eta=2$.  The other parameters are $\rho=3.5$, $\gamma=1$, $Q=0.5$, $\omega=2$ and $N=100$.}
\label{fig5}
\end{figure}

We also investigated the behavior of coupled SL oscillators in the parameter space: $(\varepsilon-Q)$ and $(\varepsilon-\eta)$. Using forward and backward continuation, we first considered the behaviour in the $(\varepsilon-Q)$ plane for $\gamma=1$, $\omega=2$, $\eta=1$, and $\rho=3.5$. This is shown in Fig.~\ref{fig4}(a). Here we observe that an increase of $Q$ leads to the {\em increase in the parameter region yielding hysteresis} (HA).  Similarly we also considered the behaviour in the $(\varepsilon-\eta)$ plane, for $\gamma=1$, $\omega=2$, $Q=0.5$, and $\rho=3.5$, shown in Fig.~\ref{fig4}(b).In this parameter space, we observe that when $\eta<1.65$, we have a wide hysteresis area, while when $\eta>1.65$, the hysteresis area suddenly decreases. 

We also display the time series of the $N$ coupled SL oscillators near the transition points. The time series of coupled systems for diffusion coefficient $\eta=1$, in both forward and backward continuation, are shown in Fig.~\ref{fig5}(a,b). In the case of forward continuation, coupled systems show synchronized oscillatory behavior before the transition at $\varepsilon=3.84$, and after the transition at $\varepsilon=3.86$, all oscillators settle to a common stable steady state, as clearly evident from Fig.~\ref{fig5}(a). In backward continuation, before the transition at  $\varepsilon=3.46$, all oscillators are in death states and after the transitions at $\varepsilon=3.44$ all oscillators show synchronized oscillatory behavior, as seen from Fig.~\ref{fig5}(b).
This synchronization of oscillations in adiabatic backward continuation of the coupling strength is expected as the states of the oscillators just before the onset of oscillatory behaviour are the same, as all oscillators are in the same death state.
The time series for the case of higher diffusion coefficient $\eta=2$ is shown in  Fig.~\ref{fig5}(c,d). Here we  can see that in forward continuation before transition at $\varepsilon=3.54$, coupled oscillators show {\em unsynchronized oscillations}, and after the transition at $\varepsilon=3.56$ the oscillators settle down to a common stable steady states (cf. Fig.~\ref{fig5}c). In backward continuation, before the transition at $\varepsilon=2.84$  all  oscillators are at the same steady state, and after the transition at $\varepsilon=2.82$, they show completely synchronized oscillation (cf. Fig.~\ref{fig5}d), as expected.  So from inspection of these time series, we clearly see that when $\eta<1.65$ the coupled systems show {\em completely synchronized oscillations} before transition and  stabilization to HSS after the transition. However when $\eta>1.65$, the coupled systems show unsynchronized oscillations before the transition and settle to HSS after the transition.

\section{Coupled Rayleigh Oscillators}

In order to gauge the generality of our results above, we now consider an ensemble of $N$ identical  Rayleigh oscillators~\cite{Ray} coupled through dynamical agents in a surrounding environment. The mathematical model of the coupled system is given by
\begin{eqnarray}
\dot{x}_i &=&\omega y_i +\varepsilon s_i \nonumber \\
\dot{y}_i &=& -\omega x_i+\rho (1-y_i^2)y_i \nonumber \\
\dot{s}_i &=& -\gamma_i s_i-\varepsilon x_i+\eta(Q\bar{s}-s_i) 
\label{eq4}
\end{eqnarray}
where $i=1,2,\ldots,N$. $\omega$ is the linear angular frequency, and $\rho>0$ governs the nonlinear friction.

\begin{figure}
\includegraphics[width=0.45\textwidth]{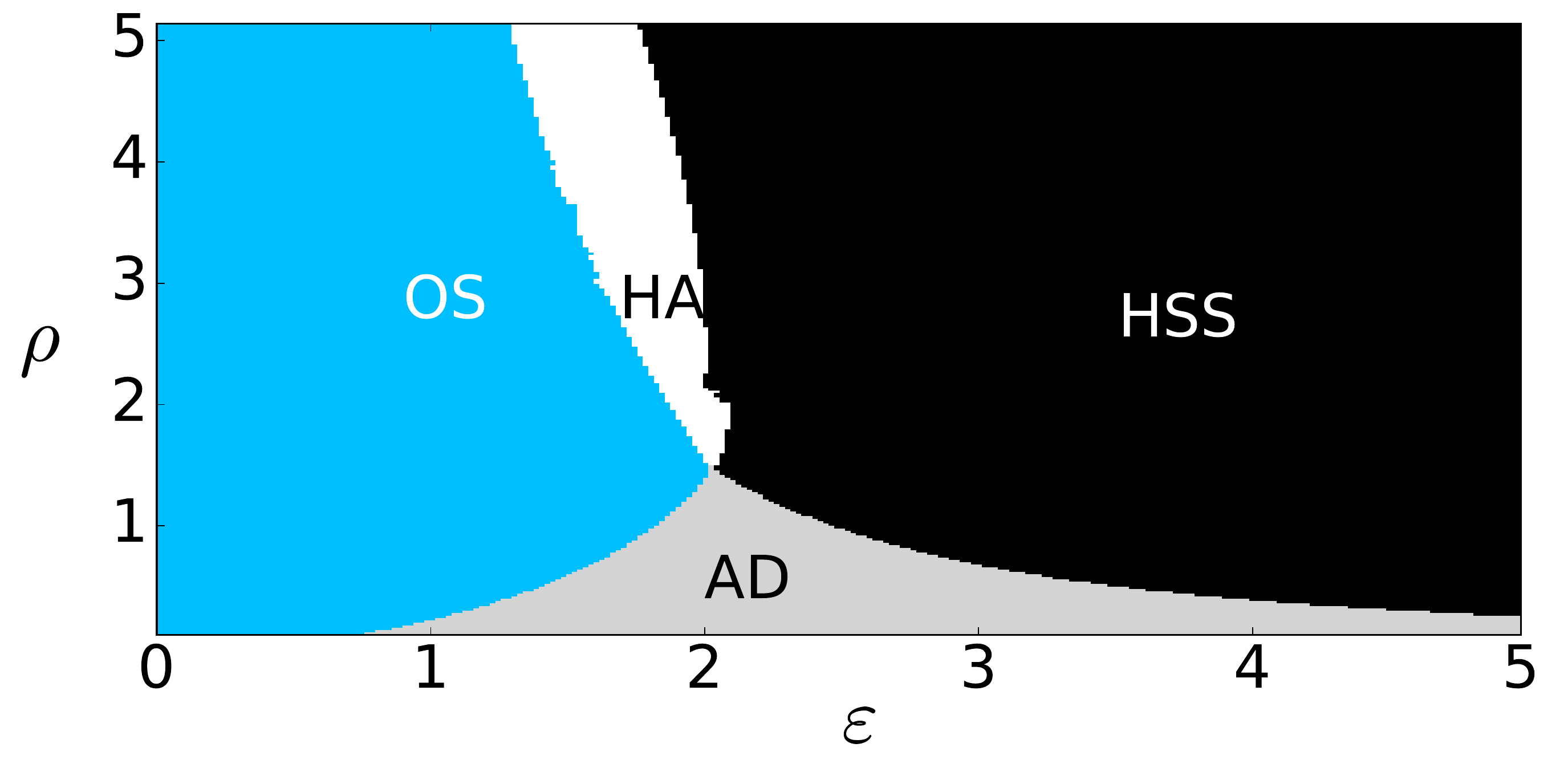}
\caption{(Color online) Phase diagram in the parameter plane $(\varepsilon -\rho)$, showing the different dynamical states arising in a system of $N$ coupled Rayleigh oscillators. The regimes marked OS, HA, HSS, and AD represent the oscillatory state, hysteresis area homogeneous steady state and amplitude death  respectively.}
\label{fig6}
\end{figure}

\begin{figure}
\includegraphics[width=0.45\textwidth]{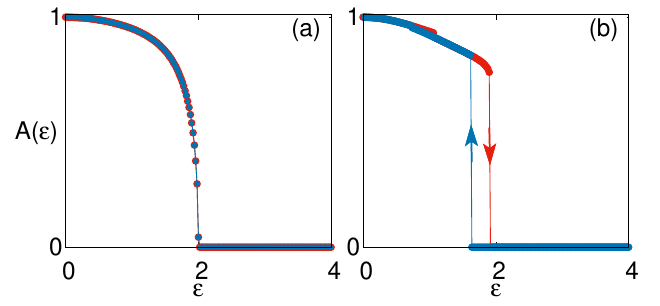}
\caption{(Color online) Forward and backward continuation of the Order Parameter, $A(\varepsilon)$, for $N=100$ coupled Rayleigh oscillators under the variation of coupling strength $\varepsilon$, are displayed for  (a) $\rho=1.5$ and (b) $\rho=4$. The other parameters are $\gamma=1$, $\omega=2$, $Q=0.5$ and $\eta=1$.}
\label{fig7}
\end{figure}

We consider $N=100$ identical  Rayleigh oscillators coupled through dynamical agents in a surrounding environment. We first calculate  a phase diagram in the parameter plane $(\varepsilon-\rho)$ adiabatically in both forward and backward continuation of $\varepsilon$. The results are shown in Fig.~\ref{fig6}, from where it is clearly evident that when $\rho<1.5$ the  coupled system stabilizes at AD, while the system stabilizes at HSS when $\rho>1.5$. The variation of  order parameters $A(\varepsilon)$  with respect to coupling strength $\varepsilon$, for both forward and backward continuation, at different values of $\rho$, are shown in Fig.~\ref{fig7}.  Fig.~\ref{fig7}(a) displays the behaviour of $A(\varepsilon)$ for $\rho=1.5$, and we see a second order transition from oscillatory state to death state and vice versa in both forward and backward continuation.  Fig.~\ref{fig7}(b) shows $A(\varepsilon)$ for $\rho=4$, and clearly indicates a discontinuous transition from oscillatory state to death state and vice versa in both forward continuation and backward continuation respectively.  So we can conclude that {\em we will observe the explosive death transition in a system of coupled Rayleigh oscillators as well, suggesting the generality of this dynamical phase transition for quorum sensing coupling}.

\begin{figure}
\includegraphics[width=0.45\textwidth]{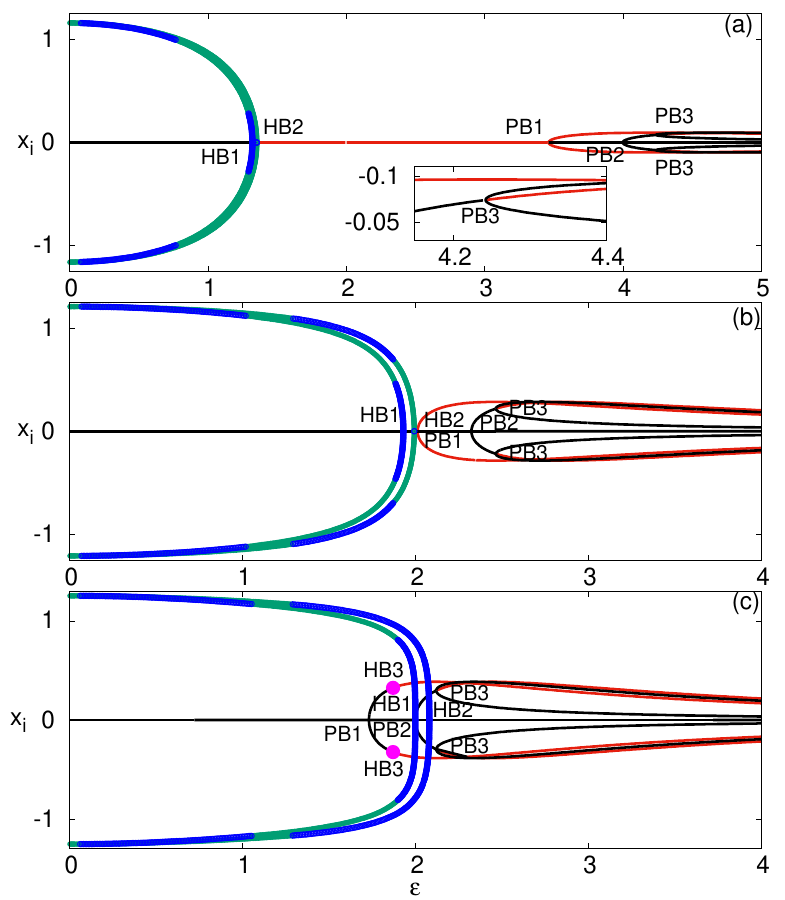}
\caption{(Color online) Bifurcation diagram of $N$ coupled Rayleigh oscillators with respect to coupling strength $\varepsilon$, for (a) $\rho=0.5$, (b) $\rho=1.5$, and (c) $\rho=2.0$. Inset of (a) shows the zoomed in view of the  pitchfork bifurcation PB3. The red and black lines show the stable and unstable steady states respectively, while the green and blue circles represent the stable and unstable periodic solutions respectively. The bifurcation points, namely pitchfork bifurcation (PB) and Hopf bifurcation (HB), are also marked. The other parameters are $\gamma=1$, $\omega=2$, $\eta=1$, $Q=0.5$ and $N=2$.}
\label{fig8}
\end{figure}

Next we show the bifurcation diagrams of two coupled Rayleigh oscillators with respect to coupling strength $\varepsilon$, for different value of $\rho$, in Fig.~\ref{fig8}. The bifurcation diagram for $\rho=0.5$ (cf. Fig.~\ref{fig8}(a)) shows that the coupled system first stabilizes at the origin through HB2, for small values of $\varepsilon$. For higher values of $\varepsilon$, the HSS solution is born and is stabilized through PB1. Further, for even higher values of $\varepsilon$, IHSS solution born through PB2 and is stabilized through another pitchfork bifurcation PB3. With increasing values of $\rho$, the bifurcation points HB2 and PB1 move closer to each other and at a critical value  $\rho_c=1.5$, HB2 collides with PB1, as shown in Fig.~\ref{fig8}(b). When $\rho>\rho_c$, HB1 and HB2 points move to the right of PB1, as seen from the bifurcation diagram for $\rho=2 >\rho_c$ in Fig.~\ref{fig8}(c). This diagram indicates that the HSS solution is born through a PB1, but stabilized through a subcritical Hopf bifurcation (HB3).

\begin{figure}
\includegraphics[width=0.45\textwidth]{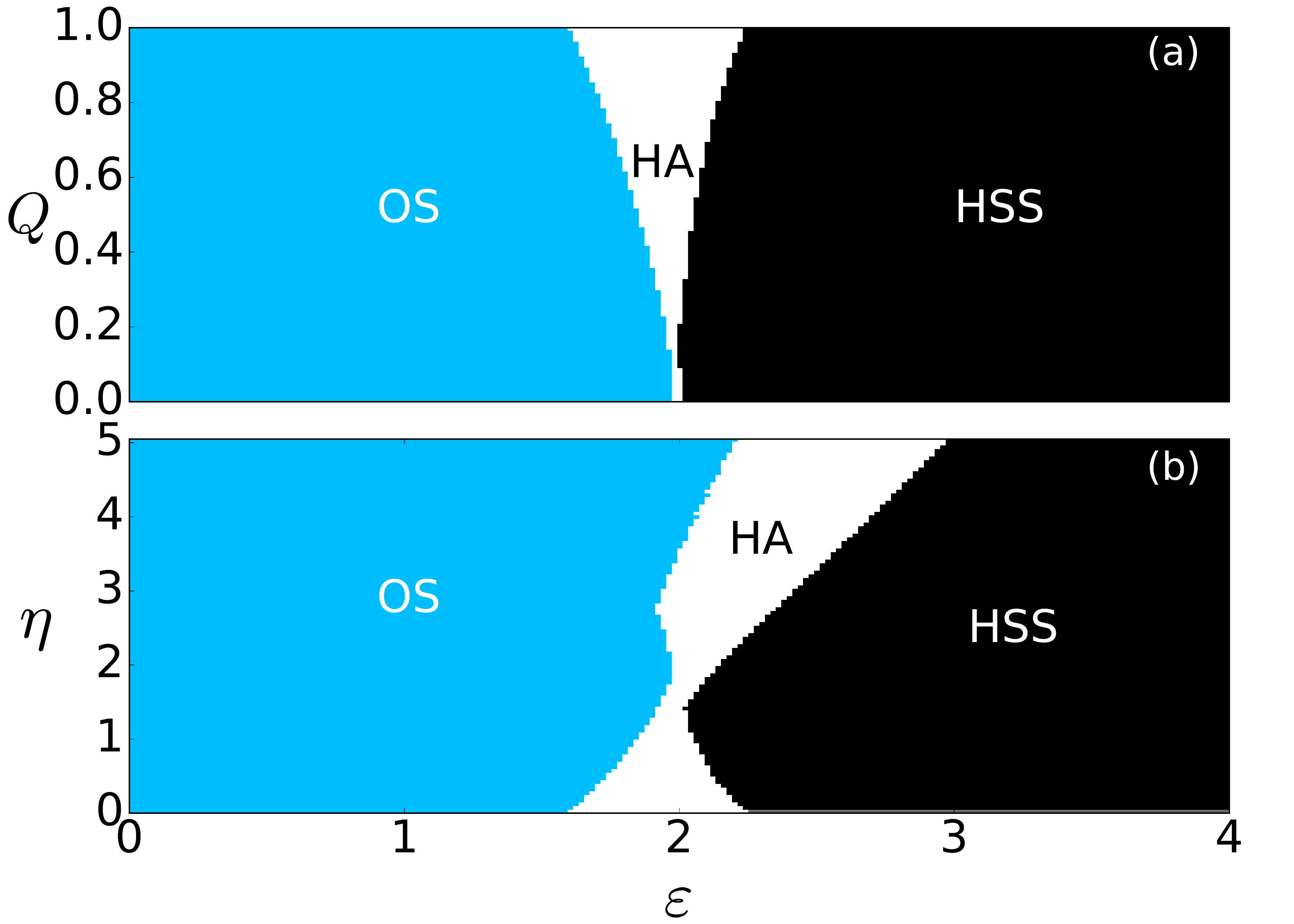}
\caption{(Color online) Different dynamical domains of coupled Rayleigh oscillators in the parameter plane (a) $(\varepsilon -Q)$ for $\eta=1$,  and (b)$(\varepsilon -\eta)$ for $Q=0.5$. The other parameters are $\rho=2$, $\gamma=1$, $\omega=2$, and $N=100$.}
\label{fig9}
\end{figure}

\begin{figure}
\includegraphics[width=0.45\textwidth]{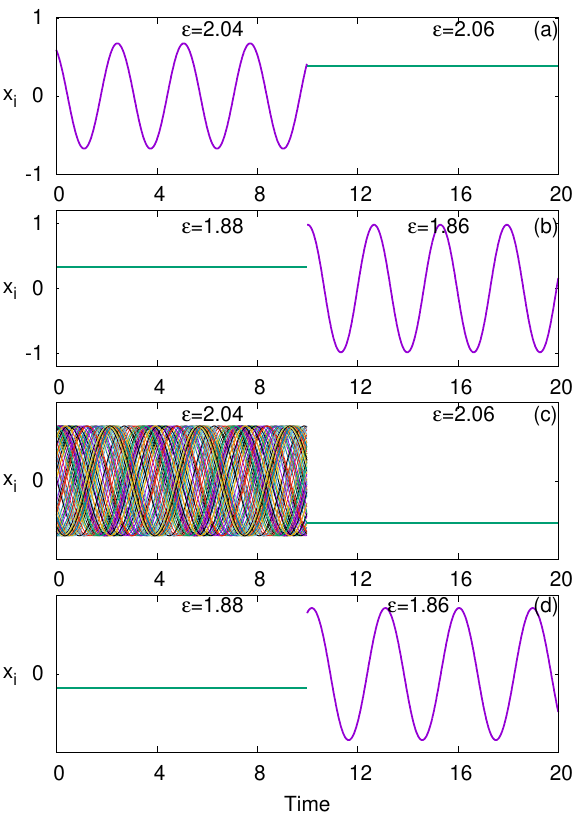}
\caption{(Color online) The time-series corresponding to Fig.~\ref{fig9}(b), near the transition points, in both forward and backward continuation 
(a,b) for $\eta=1$ and (c,d) for $\eta=2$. The other parameters are $\rho=2.0$, $\gamma=1$, $Q=0.5$, $\omega=2$ and $N=100$.}   
\label{fig10}
\end{figure}

 The behavior of coupled Rayleigh oscillators  are also studied in the parameter plane $(\varepsilon-Q)$ and $(\varepsilon-\eta)$. In Fig.~\ref{fig9}(a), the phase diagram in the parameter plane $(\varepsilon-Q)$ is shown for parameter values $\gamma=1$, $\omega=2$, $\eta=1$, and $\rho=2$. It clearly indicates that  {\em increasing $Q$ leads to the increase in the hysteresis area}. We also present the phase diagram in the parameter plane $(\varepsilon-\eta)$, for $\gamma=1$, $\omega=2$, $Q=0.5$, and $\rho=2$, in Fig.~\ref{fig9}(b). In this parameter space, we observe that when $\eta<1.4$, we have a wide hysteresis area, while when $\eta>1.4$, the hysteresis area suddenly decreases and then for higher value of $\eta$ again increases. The time series of $100$ coupled Rayleigh oscillators, for both forward and backward continuation, at $\eta=1$ are displayed in  Fig.~\ref{fig10}(a,b) respectively. They indicate that coupled systems show completely synchronized oscillations before the transition in the forward continuation and after the transition in backward continuation. Similarly, the time series, for both forward and backward continuation, at $\eta=2$ are shown in  Fig.~\ref{fig10}(c,d) respectively. These figures indicate that coupled systems show no synchronization before the transition in the forward continuation, and synchronized oscillation after the transition in backward continuation. These trends are similar to those observed for coupled Stuart-Landau oscillators, suggesting generality of the phenomena.
 
\section{Conclusion}

In summary, we have explored the spatio-temporal consequences of quorum sensing interactions by investigating model systems of nonlinear oscillators coupled via dynamical agents, from the view-point of phase transitions. Specifically, we studied the dynamics of limit cycle oscillators, namely Stuart-Landau oscillators and Rayleigh oscillators, coupled through dynamical agents who interact globally with each other in the surrounding environment. 
Interestingly, this interaction yields {\em both first-order as well as second-order transitions from the oscillatory state to death state}. In particular, for nonlinear systems with small amplitude oscillations, the coupled systems exhibit a second-order phase transition. However, for oscillators with large amplitude the coupled systems exhibit a first-order transition from oscillatory state to death state. In this first order phase transition, two states (the oscillatory and the steady state) co-exist over a range of parameter space, and this co-existence region termed a {\em hysteresis} area. We have found that  this hysteresis area and transition points of coupled systems crucially depend on the diffusive coefficient $\eta$ and  density parameter $Q$. The co-existence of oscillatory and steady state has importance in biological and chemical systems. In biological systems, the phenomena of annihilation and single-pulse triggering~\cite{bio5} can be attributed to the co-existence of these two solutions. Also in chemical systems of Paraxidase-Oxidase reaction, oscillatory solution coexists with the steady state~\cite{chem3}. So we hope that our findings of hysteresis in this model system incorporating quorum-sensing interaction will  provide more insights into the widely observed hysteresis phenomena in biological and chemical processes.


\end{document}